\documentclass[a4,10pt,twocolumn]{scrartcl}
\usepackage[top=2.5cm, bottom=2.5cm, left=1.9cm, right=1.3cm]{geometry}

\pdfoutput=1

\usepackage[english]{babel}
\usepackage[latin1]{inputenc}

\usepackage{graphicx} % for pdf, bitmapped graphics files
\usepackage{epsfig} % for postscript graphics files
\usepackage{libertine}
\usepackage[T1]{fontenc}
\usepackage{mathptmx} % assumes new font selection scheme installed
\usepackage{amsmath} % assumes amsmath package installed
\usepackage{amssymb}  % assumes amsmath package installed
\usepackage{mathdots}
\usepackage{xcolor}

\usepackage{framed}
\colorlet{shadecolor}{green!60!black!0} 	% Farbe für Boxen

\usepackage[hang]{footmisc}
\setfnsymbol{wiley}

\makeatletter
\renewcommand{\@biblabel}[1]{[#1]\hfill}
\makeatother

\usepackage{amsthm}

\usepackage{tikz}				
\usetikzlibrary{arrows}				
\usetikzlibrary{shapes.misc}
\usetikzlibrary{patterns}
\usepackage{pgfplots}
\usetikzlibrary{calc}
\usepackage{setspace}

\usepackage{siunitx}

\makeatletter
\let\NAT@parse\undefined
\makeatother

\colorlet{lcolor}{green!50!black}
\colorlet{ucolor}{magenta!40!black}
\colorlet{ccolor}{green!40!black}

\usepackage[colorlinks=true,%
            linkcolor=lcolor,%
            urlcolor=lcolor,%
            citecolor=lcolor]{hyperref}

\usepackage[nameinlink]{cleveref}

\usepackage[nomessages]{fp}

\usepackage[format=plain,			% no indentation in second line of caption
            labelsep=period,		% period after ``Figure''
            font=small,			% set font size
            labelfont=bf,			% bold labels
            skip=5pt			% spacing between caption and figure
            ]{caption}
            
\allowdisplaybreaks[1]			% allows pagebreaking in align
\renewenvironment{thebibliography}[1]{%
       \begin{oldthebibliography}{#1}%
       \setlength{\parskip}{0ex}%
       \setlength{\itemsep}{0ex}%
}%
{%
        \end{oldthebibliography}%
}

\usepackage[final,kerning=false,protrusion=false]{microtype}

\newtheorem{theorem}{Theorem }
\newtheorem{lemma}{Lemma }
\newtheorem{corollary}{Corollary }
\newtheorem{probState}{Main Problem }
\newtheorem{definition}{Definition }
\newtheorem{remark}{Remark }

\let\oldTheorem\theorem
\renewcommand{\theorem}{\oldTheorem\normalfont}
\let\oldLemma\lemma
\renewcommand{\lemma}{\oldLemma\normalfont}
\let\oldCorollary\corollary
\renewcommand{\corollary}{\oldCorollary\normalfont}
\let\oldProbState\probState
\renewcommand{\probState}{\oldProbState\normalfont}
\let\oldDefinition\definition
\renewcommand{\definition}{\oldDefinition\normalfont}
\let\oldRemark\remark
\renewcommand{\remark}{\oldRemark\normalfont}

\newcommand\myComment[1]{\textsc{\textcolor{red!60!black}{#1}}}
\renewcommand\myComment[1]{}			% uncomment this line to hide todo-Notes

\title{\LARGE \bf
Gradient approximation and extremum seeking \\ via needle variations 
\footnote{This article is an extended version of \cite{mic2016needle}.}
}

\setkomafont{section}{\Large}
\setkomafont{subsection}{\large}
\setkomafont{subsubsection}{\normal}

\begin{document}
% \setlength{\abovedisplayshortskip}{1.35ex plus1ex minus1ex}
% \setlength{\abovedisplayskip}{1.35ex plus1ex minus1ex}
% \setlength{\belowdisplayshortskip}{1.35ex plus1ex minus1ex}
% \setlength{\belowdisplayskip}{1.35ex plus1ex minus1ex}

%\date{}
\author{Simon Michalowsky and Christian Ebenbauer % <-this % stops a space
%\thanks{This work was supported by Deutsche Forschungsgemeinschaft (Emmy-Noether-Grant, ????)}% <-this % stops a space
\thanks{S. Michalowsky and C. Ebenbauer are with the Institute for Systems Theory and Automatic Control,
University of Stuttgart, 70550 Stuttgart, Germany.
{\tt\small simon.michalowsky@ist.uni-stuttgart.de, ce@ist.uni-stuttgart.de}.
This work was supported by the German Research Foundation (DFG, EB 42S/4-1).} %
}

% \begin{titlepage}
% \begin{center}
% {\LARGE \bf Gradient approximation and extremum seeking via needle variations} \\[2em]
% {\Large Simon Michalowsky and Christian Ebenbauer}
% }
% \end{center}
% \newpage
% \end{titlepage}

\maketitle

% \twocolumn[
%   \begin{@twocolumnfalse}
%     \maketitle
%     This article is an extended version of (Referenz auf ACC-Paper).
%     Most of the content is equal but it contains additional material
%     in the appendix. \par\vspace*{2em}
%   \end{@twocolumnfalse}
% ]

% \maketitle
% \thispagestyle{empty}
% \pagestyle{empty}

%%%%%%%%%%%%%%%%%%%%%%%%%%%%%%%%%%%%%%%%%%%%%%%%%%%%%%%%%%%%%%%%%%%%%%%%%%%%%%%%
\begin{abstract}
\textbf{Abstract.} We consider a gradient approximation scheme that is based on applying needle
shaped inputs. By using ideas known from the classic proof of the Pontryagin
Maximum Principle we derive an approximation that reveals that the considered
system moves along a weighted averaged gradient. Moreover, based on the same 
ideas, we give similar results for arbitrary periodic inputs. We also present
a new gradient-based optimization algorithm that is motivated by our
calculations and that can be interpreted as a combination of the heavy ball
method and Nesterov's method. 
\end{abstract}

%%%%%%%%%%%%%%%%%%%%%%%%%%%%%%%%%%%%%%%%%%%%%%%%%%%%%%%%%%%%%%%%%%%%%%%%%%%%%%%%
\section{Introduction}\label{secIntroduction}
% A large class of problems admits a gradient-based solution, especially
% in cases where some optimality criterion is involved. However, in many
% of these problems the gradient cannot be computed and a common approach
% is to use gradient approximation schemes. One of these schemes is
% extremum seeking where the goal is to operate a dynamical system at an
% optimal steady state. Since the steady state map and often also the 
% performance function are commonly unknown, a gradient-based scheme 
% cannot be used and
Extremum seeking is a well-known technique that has successfully been
used in several applications (see e.g. \cite{cougnon2011line}, \cite{wang2000experimental}, \cite{zhang2011extremum}) in order to operate a 
system at an a priori unknown setpoint that is optimal with respect to some objective function.
In a typical extremum seeking problem this objective
function is unknown such that gradient-based methods do not apply.
The majority of the extremum seeking schemes relies on approximating the gradient % of a steady-state map of a dynamic system
by the excitation of the system with a so called dither signal. Although a
much larger class of periodic signals is appropriate in principle, in most cases
sinusoidal dither signals are used. 
% It is shown in \cite{tan2008dither} that
% this common choice of dither signals might not always be the best choice.
% 
% In most cases sinusoidal
% dither signals are used although a much larger class of periodic signals
% is appropriate in principle and despite the fact that 
%
The effect of the choice of the dither has been studied
in \cite{tan2008dither} where the authors conclude that dither signals
other than sinusoidal ones could be beneficial. \\
In the present paper we introduce periodic needle-shaped dither signals for gradient
approximation. In particular we consider the case where the period length is large compared to
the pulse length. This is in contrast to standard averaging results (\cite{krstic2003real})
or Lie bracket averaging results (\cite{duerr2013LieBracket}) where the period length is
also assumed to be small. 
Our main objective is not to present a new extremum seeking scheme here but
to get insight into the extremum seeking process using needle-shaped dither signals. Nevertheless, this class of dithers
could also be of interest in certain applications, e.g. when it is only possible
to apply inputs in a short period of time. \\
Our contributions are as follows:
For the case of needle-shaped dither signals we show that the extremum seeking
system approximates a weighted averaged gradient descent. 
% Our results recover this standard case in the limit, but 
% they further allow us to get more insight into the approximation process and reveal
% that a weighted averaged gradient is approximated. \\
% One of our motivations to this is to get more
% insight into the approximation process of extremum seeking schemes: While standard averaging
% results (\cite{krstic2003real}) or Lie bracket averaging results (\cite{duerr2013LieBracket})
% tell us that in the limit, i.e. for decreasing period length,
% a gradient is approximated, the error introduced when not looking at
% the limit is not yet sufficiently characterized. 
Secondly, we use that a large class of dither signals 
is well approximated by a summation of multiple needle-shaped signals and we show how
extremum seeking systems with such dither signals can be analyzed with the presented 
theory as well. We establish new results for finite period length that are in the
limit in accordance with existing results. 
We further propose a new gradient-based optimization algorithm
that is indeed motivated by the results on gradient approximation via needle variations
but can also be regarded separately. The algorithm can be seen as a combination of
a continuous-time version of Nesterov's method and the heavy ball method and is of interest on its own. \\
Our analysis relies on well-known results established in the context of the Pontryagin
Maximum Principle where the effects of needle-shaped variations
of the optimal inputs are studied. Since the Maximum Principle
is well-established in a very broad setup we hope that our
ideas can be used for much more general problems. However, our
results should be seen as a first step and in the present paper we only 
consider basic cases. \\
The structure of the present paper is as following: In \Cref{secPreliminaries}
we give a brief introduction to needle variations and the variational
equations. In \Cref{secMainResults} we present our main results. We 
first consider the case of two needles with opposite sign, then we show how
this can be generalized to a superposition of many needles and third we
present the new continuous-time algorithm. In \Cref{secExample} we illustrate our approximation
formulas for the particular case of a quadratic objective function and
give simulation results for the proposed new algorithm. We conclude our
work in \Cref{secConclusions}.

%%%%%%%%%%%%%%%%%%%%%%%%%%%%%%%%%%%%%%%%%%%%%%%%%%%%%%%%%%%%%%%%%%%%%%%%%%%%%%%%
\section{Preliminaries}\label{secPreliminaries}
\subsection{Notation}
We denote by $ \mathbb{N} $ the set of natural numbers, by $ \mathbb{Z} $ the set of
integer numbers and by $ \mathbb{R} $ the set of real numbers.
We denote by $ \mathcal{C}^n $ the set of $n$ times continuously differentiable functions.
The gradient of a function \mbox{$ f: \mathbb{R}^p \to \mathbb{R} $}, \mbox{$ f \in \mathcal{C}^1 $}
is denoted by
%$
%     \nabla f(x) = \begin{bmatrix} \frac{\partial f}{\partial x_1}(x) & \frac{\partial f}{\partial x_2}(x) & ... & \frac{\partial f}{\partial x_p}(x) \end{bmatrix}^\top .
%$
$
     \nabla f(x) = [ \frac{\partial f}{\partial x_1}(x) \quad \frac{\partial f}{\partial x_2}(x) \quad ... \quad \frac{\partial f}{\partial x_p}(x) ]^\top .
$
Moreover, we make use of the Landau notation: For $ f,g: \mathbb{R}^n \to \mathbb{R} $
we write
$
	f(x) = \mathcal{O}\big( g(x) \big)
$
meaning that there exists some $ M>0 $ and some $\delta>0$ such that
$
	\vert f(x) \vert \leq M \vert g(x) \vert 
$
for all $ \vert x \vert \leq \delta $.

\subsection{Needle variations and the variational equation}\label{subsectionPrelimVariationalEquation}
\begin{figure}[t]
	\vspace*{0.2cm}
	\begin{center}
		\begin{tikzpicture}[>=latex]
	%%%%%%%%%%%%%%%%%%%%%%%%%%%%%%%%%%%%%%%%%%%%%%%%%%%%%%%%%%%%%%%%%%%%%%%%%%%%%%%%%%%%%%%%%%%%%%%%%%%%%%%%%%%%%
	% N E E D L E   V A R I A T I O N S   O F   T H E   I N P U T S
	% draw coordinate axis
	\draw[->] (0,0) -- (2.5,0);
	\draw[->] (0,0) -- (0,2.6);
	% draw variation
	\draw[very thick,green!60!black] (-0.1,0.01) -- (1,0.01);
	\draw[green!60!black,dashed] (1,0.01) --  (1,2.3);
	\draw[very thick,green!60!black] (1,2.3) -- (1.3,2.3);
	\draw[green!60!black,dashed] (1.3,2.3) -- (1.3,0.01);
	\draw[very thick,green!60!black] (1.3,0.01) -- (2.3,0.01); 
	% draw tick at alpha
	\draw[] (-0.1,2.3) -- node[pos=1,anchor=west] {$\alpha$} (0.1,2.3);
	% label variation
	\node[anchor=west,green!60!black] at (1.3,2.2) {$ u_2(t) $};
	% draw optimal u_2^* = 0
	\draw[dashed,ultra thick,orange] (-0.1,0.01) -- (2.3,0.01);
	% label optimal u_2
	\node[orange,anchor=south] at (2,0.01) {$ u_2^*(t) $};
	% draw tick at needle
	\draw[] (1,0.1) -- node[pos=1,anchor=north east,xshift=0.5em] {$\bar{t}-\varepsilon$} (1,-0.1);
	\draw[] (1.3,0.1) -- node[pos=1,anchor=north] {$\bar{t}$} (1.3,-0.1);
	%%%%%%%%%%%%%%%%%%%%%%%%%%%%%%%%%%%%%%%%%%%%%%%%%%%%%%%%%%%%%%%%%%%%%%%%%%%%%%%%%%%%%%%%%%%%%%%%%%%%%%%%%%%%%
	% E F F E C T S    O N    T H E    T R A J E C T O R I E S
	% draw starting point
	\node[name=x0,draw,circle,fill=black,inner sep=0pt,minimum width=4pt,label=135:$x_0$] at (3.8,0.3) {};
	% get point where the needle starts
	\path[] (x0) to [out=-10,in=-155] node[draw=black,circle,fill=black,inner sep=0pt,minimum width=4pt,pos=1,name=startNeedle] {} ++(1.5,0.7) {};
	\node[anchor=south east] at (startNeedle) {$x^*(\bar{t}-\varepsilon)$};
	% draw trajectory after needle
	\draw[green!60!black,thick] (x0) to [out=-10,in=-155] (startNeedle.center) 
	                                 to [bend right=15] node[pos=1,fill=black,name=endNeedle,draw=black,circle,inner sep=0pt,minimum width=4pt] {} ++(0.3,0.7) 
	                                 to [out=180-157,in=-173] node[anchor=south west] {$x(t)$} ++(2,0.3) ;
	\draw[green!60!black,thick] (x0) to [out=-10,in=-155] (startNeedle.center) 
	                                 to [bend right=15] ++(0.3,0.7) 
	                                 to [out=180-157,in=-173] node[anchor=south west] {$x(t)$} ++(2,0.3) ;
	\node[anchor=south east] at (endNeedle) {$x(\bar{t})$};
	% draw optimal trajectory
	\draw[orange,dashed,thick] (x0) to [out=-10,in=-155] (startNeedle)  
	                                to [out=180-155,in=-170] node[anchor=north west] {$x^*(t)$} ++(2.5,0.3);
	
\end{tikzpicture}
	\end{center}
	\caption{Illustration of the needle variation as defined by \eqref{eqPrelimNeedleVariation} 
	         (left) as well as the optimal and the varied trajectory (right).}
	\label{figPrelimNeedleVariation}
\end{figure}
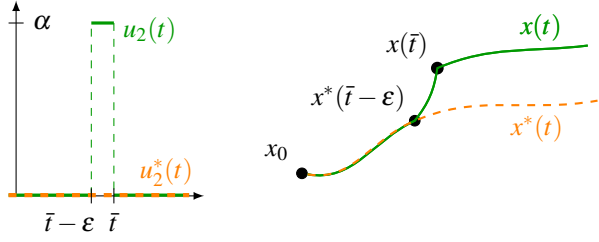

In the following we briefly repeat well-known results on the effect of
so called needle variations in the control inputs to trajectories of
dynamic systems. Our overview follows closely the lines of \cite[Chapter 4]{liberzon2011calculus}
but is adapted to the special needs for the argumentation of our main results.
Consider
\begin{align}
	\dot{x}(t) = g_1\big(x(t)\big) u_1(t) + g_2\big(x(t)\big) u_2(t)  \label{eqPrelimInputAffineSystem}
\end{align}
where $ g_1, g_2 : \mathbb{R} \to \mathbb{R} $ and $ u_1, u_2 : \mathbb{R} \to \mathbb{R} $.
Further, suppose that $ g_1, g_2 \in \mathcal{C}^1 $ and $ u_1, u_2 $ are
piecewise continuous such that at least local existence and uniqueness of
a solution to \eqref{eqPrelimInputAffineSystem} is ensured.
\myComment{(What is required for the vector fields? We later need that the
solution is analytic for the Taylor expansion.)}
Let $ x^*(t) $ denote the solution of \eqref{eqPrelimInputAffineSystem} for
$ u_1(t) = u_1^*(t) $ and $ u_2(t) = u_2^*(t) \equiv 0 $. We are interested in how the solution
$ x^*(t) $ will change when we perturb the input $ u_2^* $ by a so called
needle variation, also known as Pontryagin-McShane variation, where the
perturbed input is defined as
\begin{align}
	u_2(t) = \begin{cases}
			u_2^*(t)  & \text{if } t \notin [\bar{t}-\varepsilon,\bar{t}] \\
			\alpha    & \text{if } t \in [\bar{t}-\varepsilon,\bar{t}]
		\end{cases}
		\label{eqPrelimNeedleVariation}
\end{align}
with some constant $ \alpha \in \mathbb{R} $ and $ \varepsilon > 0 $. Thus,
the input is perturbed on an interval of length $ \varepsilon $ by some
constant value $ \alpha $, see \Cref{figPrelimNeedleVariation}. In the following
we will investigate the effect of such a perturbation for small $ \varepsilon $.
Let $ x(t) $ denote the solution of \eqref{eqPrelimInputAffineSystem} when applying the perturbed
input and suppose that $ u_1^*(t) $ is continuous at $ t = \bar{t} $. % $ \bar{t} $ is a point of continuity of $ u_1^*(t) $.
\myComment{(Is this really needed? Liberzon states this is required for the Taylor expansion. - This is sufficient for $ x^*(t) $ be continuously differentiable at $\bar{t} $ which is required for the Taylor expansion up to first order terms.)}
By some Taylor expansions one can show that (see e.g. \cite{liberzon2011calculus})
\begin{align}
	x(\bar{t}) &= x^*(\bar{t}) + \varepsilon \big[ g_1\big(x^*(\bar{t})\big) u_1^*(\bar{t}) + g_2\big(x^*(\bar{t})\big) \alpha \nonumber \\
	              &\phantom{= x^*(\bar{t}) + \varepsilon \big[}- g_1\big(x^*(\bar{t})\big) u_1^*(\bar{t})  \big] + \mathcal{O}(\varepsilon^2) \nonumber \\
	           &= x^*(\bar{t}) + \varepsilon g_2\big(x^*(\bar{t})\big) \alpha + \mathcal{O}(\varepsilon^2).
	           \label{eqPrelimAfterNeedle}
\end{align}
We will now investigate how the perturbed solution $ x(t) $ is propagated
after time $ \bar{t} $ in comparison to the unperturbed solution $ x^*(t) $.
This can be studied using perturbation theory (\cite{khalil2002nonlinear}) where one is interested in how 
the solution of a differential equation evolves when starting from a perturbed
initial condition compared to the solution when starting from a nominal
initial condition. Here, $ x^*(\bar{t}) $ plays the role of the nominal
initial condition whereas the perturbed one is given by \eqref{eqPrelimAfterNeedle}.
The basic idea is to do a Taylor expansion of the perturbed solution in $ \varepsilon $
about $ \varepsilon = 0 $ which leads to
\begin{align}
	x(t) = x^*(t) + \varepsilon v_1(t) + \mathcal{O}(\varepsilon^2) \label{eqPrelimTaylorPerturbed}
\end{align}
where $ v_1(t) $ is some unknown function that can be determined as following. 
Since $ x(t) $ must fulfill the unperturbed differential equation one can put
$ x(t) $ as given by \eqref{eqPrelimTaylorPerturbed} into \eqref{eqPrelimInputAffineSystem}
with $ u_1 = u_1^* $, $ u_2 = u_2^* $ and compare the terms linear in $ \varepsilon $
which then gives (\cite{liberzon2011calculus},\cite{khalil2002nonlinear})
\begin{align}
	\dot{v}_1(t) = \tfrac{\partial g_1}{\partial x}\big( x^*(t) \big) u_1^*(t) v_1(t) \label{eqPrelimVariationalEquation}
\end{align}
with $ v_1(\bar{t}) = g_2\big(x^*(\bar{t})\big) \alpha $. This is known as
the variational equation (\cite{liberzon2011calculus}) and it is the same as
the linearization of \eqref{eqPrelimInputAffineSystem} with $ u_1(t) = u_1^*(t) $,
$ u_2(t) = u_2^*(t) $ about the trajectory $ x^*(t) $.

%\afterpage{\clearpage}

%%%%%%%%%%%%%%%%%%%%%%%%%%%%%%%%%%%%%%%%%%%%%%%%%%%%%%%%%%%%%%%%%%%%%%%%%%%%%%%%
\section{Main results}\label{secMainResults}
We consider the following nonlinear input-affine system
\begin{align}
	\dot{x}(t) = F\big(x(t)\big) u_1(t) + u_2(t)                               \label{eqNonlinearInputAffineSys}
\end{align}
with $ x(0) = x_0 \in \mathbb{R} $ and where $ F: \mathbb{R} \to \mathbb{R} $,
$ F \in \mathcal{C}^1 $. In extremum seeking problems the usual approach is to choose the inputs
$ u_1, u_2 $ such that the trajectories of \eqref{eqNonlinearInputAffineSys} approximately
move along those of the gradient flow
$
	\dot{\bar{x}}(t) = -\nabla F\big( \bar{x}(t) \big),
$
$ \bar{x}(0) = x_0 $, such that $x$ converges to a solution of the
optimization problem $ \min \, F(x) $ (see e.g. \cite{krstic2003real} or \cite{duerr2013LieBracket}).
Here, we want to investigate the averaged behavior of \eqref{eqNonlinearInputAffineSys} for a new class
of inputs: 
We first consider needle-shaped inputs and then give a generalization thereof.
We consider the scalar case here such that, strictly speaking, we are only
treating derivative approximation. However it should be noted that 
one can expect similar results for the multidimensional case.

%%%%%%%%%%%%%%%%%%%%%%%%%%%%%%%%%%%%%%%%%%%%%%%%%%%%
\subsection{Dither signals composed of two needles}\label{subsectionTwoNeedles}
Define the following $ T $-periodic input sequence 
\begin{align}
	u_1(t) &= \begin{cases}
				1  & \text{for } t \in [0, \tfrac{T}{2}) \\
				-1 & \text{for } t \in [\tfrac{T}{2},T) 
	         \end{cases}  \label{eqInputSequence1_u1} \\
	u_2(t) &= \begin{cases}
				\alpha  & \text{for } t \in [0,\varepsilon) \\
				0       & \text{for } t \in [\varepsilon,\tfrac{T}{2}) \\
				-\alpha & \text{for } t \in [\tfrac{T}{2}, \tfrac{T}{2} + \varepsilon ) \\
				0       & \text{for } t \in [\tfrac{T}{2} + \varepsilon,T) 
	         \end{cases} \label{eqInputSequence1_u2}                     
\end{align}
where $ T >0 $, $ 0 < \varepsilon < \tfrac{T}{2} $ and $ -\infty < \alpha < \infty $. The input sequence is illustrated
in \Cref{figInputSequenceTwoNeedles}. The following theorem reveals how this input sequence affects \eqref{eqNonlinearInputAffineSys}
for $ \varepsilon $ being small.
\begin{snugshade}
\begin{theorem}\label{theoremTwoNeedles}
	Consider \eqref{eqNonlinearInputAffineSys} together with the input sequence
	as defined by \eqref{eqInputSequence1_u1} and \eqref{eqInputSequence1_u2}.
	Let $ x^*(t) $ denote the solution of \eqref{eqNonlinearInputAffineSys} when
	$ u_1(t) $ as defined by \eqref{eqInputSequence1_u1} and $ u_2(t) \equiv 0 $
	and suppose that $ x^*(t) $ exists on $ [0,T] $.
	Let $ \Phi(t,t_0) $ denote the state-transition matrix at time $t$ corresponding
	to the variational equation 
	\begin{align}
		\dot{v}_1(t) = \tfrac{\partial F}{\partial x} \big( x^*(t) \big) v_1(t) \label{eqTwoNeedlesVariationalEquation}
	\end{align}
	with initial time $t_0$. Then
	\begin{align}
		&x(T) = \label{eqApproximationTwoNeedles1} \\ & x_0 + \varepsilon \alpha \Phi(0,\varepsilon) \int_{\varepsilon}^{\tfrac{T}{2}-\varepsilon} \tfrac{\partial F}{\partial x} \big( x^*(\tau) \big) \Phi(\tau,\tfrac{T}{2}-\varepsilon) \, d\tau + \mathcal{O}(\varepsilon^2).  \nonumber
	\end{align}
\end{theorem}
\end{snugshade}
\def\figurewidth{\columnwidth}
\def\figureheight{4cm}
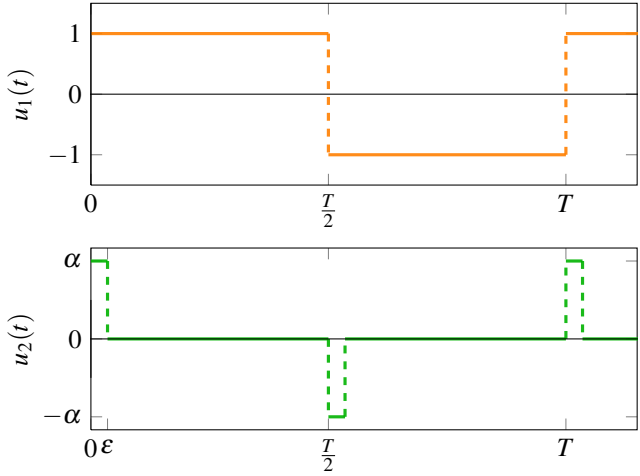
\begin{figure}[t]
	\vspace*{-1em}   % this vspace should only be used if the figure is aligned on top of the page in order to fit the page margins
	\def\eps{0.07} 
\colorlet{u1Color}{orange}
\begin{tikzpicture}[>=latex]
	\begin{axis}[
			width=\figurewidth,
			height=\figureheight,
			xmin=0, 
			xmax=2.3,
			ymin=-1.5, 
			ymax=1.5,
			disabledatascaling,
			xlabel=$ $,
			ylabel=$u_1(t)$,
			xtick={0,1,2},
			xticklabels={ $0$,$\tfrac{T}{2}$,$T$  },
			ytick={-3,-1,0,1,3},
			yticklabels={$-\alpha$,$-1$,$0$,$1$,$\alpha$},
			ylabel style={yshift=-0.3cm},
			legend style={at={(0.55,0.97)},anchor=north west,nodes=right},
			grid=none]
		% plot u2
		\addplot[very thick,color=u1Color,opacity=0.9] coordinates {(0,1) (1,1)};
		\addplot[very thick,color=u1Color,opacity=0.9,dashed] coordinates {(1,1) (1,-1)};
		\addplot[very thick,color=u1Color,opacity=0.9] coordinates {(1,-1) (2,-1)};
		\addplot[very thick,color=u1Color,opacity=0.9,dashed] coordinates {(2,-1) (2,1)};
		\addplot[very thick,color=u1Color,opacity=0.9] coordinates {(2,1) (3,1)};
		% coordinate axis
		\draw[-] (axis cs: -1,0) -- (axis cs: 6,0);
		\draw[-] (axis cs: 0,-1.5) -- (axis cs: 0,1.5);
	\end{axis}
\end{tikzpicture}
\begin{tikzpicture}
	\begin{axis}[
			width=\figurewidth,
			height=\figureheight,
			xmin=0, 
			xmax=2.3,
			ymin=-3.5, 
			ymax=3.5,
			disabledatascaling,
			xlabel=$ $,
			ylabel=$u_2(t)$,
			xtick={0,\eps,1,2},
			xticklabels={ $0$,$\varepsilon$,$\tfrac{T}{2}$,$T$  },
			ytick={-3,0,3},
			yticklabels={$-\alpha$,$0$,$\alpha$},
			ylabel style={yshift=-0.3cm},
			legend style={at={(0.55,0.97)},anchor=north west,nodes=right},
			grid=none]
		% plot u1
		\addplot[very thick,color=green!70!black,opacity=0.9] coordinates {(0,3) (\eps,3)};
		\addplot[very thick,color=green!70!black,opacity=0.9,dashed] coordinates {(\eps,3) (\eps,0)};
		\addplot[very thick,color=green!70!black,opacity=0.9] coordinates {(\eps,0) (1,0)};
		\addplot[very thick,color=green!70!black,opacity=0.9,dashed] coordinates {(1,0) (1,-3)};
		\addplot[very thick,color=green!70!black,opacity=0.9] coordinates {(1,-3) (1+\eps,-3)};
		\addplot[very thick,color=green!70!black,opacity=0.9,dashed] coordinates {(1+\eps,-3) (1+\eps,0)};
		\addplot[very thick,color=green!70!black,opacity=0.9] coordinates {(1+\eps,0) (2,0)};
		\addplot[very thick,color=green!70!black,opacity=0.9,dashed] coordinates {(2,0) (2,3)};
		\addplot[very thick,color=green!70!black,opacity=0.9] coordinates {(2,3) (2+\eps,3)};
		\addplot[very thick,color=green!70!black,opacity=0.9,dashed] coordinates {(2+\eps,3) (2+\eps,0)};
		\addplot[very thick,color=green!70!black,opacity=0.9] coordinates {(2+\eps,0) (3,0)};
		% coordinate axis
		\draw[-] (axis cs: -1,0) -- (axis cs: 6,0);
		\draw[-] (axis cs: 0,-1.5) -- (axis cs: 0,1.5);
	\end{axis}
\end{tikzpicture}
	\caption{Illustration of the input sequence as defined by \eqref{eqInputSequence1_u1} and \eqref{eqInputSequence1_u2}.
	         The upper plot shows $ u_1(t) $, the lower one shows the needle variation $ u_2(t) $.}
	\label{figInputSequenceTwoNeedles}
\end{figure}
\begin{proof}
	We use the results presented in \Cref{subsectionPrelimVariationalEquation}
	where $ u_2 $ plays the role of the needle variation. With \eqref{eqPrelimAfterNeedle}
	we have after the first needle, i.e. at $ t = \varepsilon $, 
	$
		x(\varepsilon) = x^*(\varepsilon) + \varepsilon \alpha + \mathcal{O}( \varepsilon^2 ).
	$
	Using \eqref{eqPrelimTaylorPerturbed} we have at the point where
	the second needle is applied, i.e. at $ t = \tfrac{T}{2} $,
	\begin{align}
		x(\tfrac{T}{2}) = x^*(\tfrac{T}{2}) + \varepsilon v_1(t) + \mathcal{O}( \varepsilon^2 ) \label{eqProofTwoNeedlesBeforeSecondNeedle}
	\end{align}
	where $ v_1(t) $ is the solution of \eqref{eqTwoNeedlesVariationalEquation}
	with initial condition \mbox{$ v_1(\varepsilon) = \alpha $}. Since $ \Phi(t,t_0) $
	denotes the state-transition matrix corresponding to \eqref{eqTwoNeedlesVariationalEquation}
	we have that
	\mbox{%
	$
		v_1(t) = \Phi(t,\varepsilon) v_1(\varepsilon) = \Phi(t,\varepsilon) \alpha
	$}
	for $ t \in [\varepsilon, \tfrac{T}{2}] $ such that
	\begin{align}
		x(\tfrac{T}{2}) = x^*(\tfrac{T}{2}) + \varepsilon \Phi(\tfrac{T}{2},\varepsilon) \alpha + \mathcal{O}( \varepsilon^2 ) . \label{eqrefProofTwoNeedlesBeforeSecondNeedle2}
	\end{align}
	We will now repeat the same procedure for the second needle. Let $ \bar{x}(t) $
	denote the solution when only applying the first but not the second needle, 
	i.e. we have
	\mbox{$
		\bar{x}(t) = x^*(t) + \varepsilon \bar{\Phi}(t,\varepsilon) \alpha + \mathcal{O}( \varepsilon^2 )
	$}
	where $ \bar{\Phi}(t,t_0) $ denotes the state-transition matrix corresponding to
	\begin{align}
		\dot{v}(t) = \tfrac{\partial F}{\partial x}\big( x^*(t) \big) u_1(t) v(t) \label{eqProofTwoNeedlesVarEq2}
	\end{align}
	with initial time $ t_0 $. In comparison to the previous calculations $ \bar{x}(t) $
	now plays the role of $ x^*(t) $ in \eqref{eqPrelimAfterNeedle} such that
	\begin{align}
		x(\tfrac{T}{2}+\varepsilon) = \bar{x}( \tfrac{T}{2}+\varepsilon ) - \varepsilon \alpha + \mathcal{O}(\varepsilon^2).
	\end{align}
	Hence, propagating the second needle, we obtain by \eqref{eqPrelimTaylorPerturbed} 
	\begin{align}
		x(t) &= \bar{x}(t) + \varepsilon {v}_2(t) + \mathcal{O}(\varepsilon^2) \nonumber \\
		     &= x^*(t) + \varepsilon \big( \bar{\Phi}(t,\varepsilon) \alpha + {v}_2(t) \big) + \mathcal{O}(\varepsilon^2) 
	\end{align}
	for $ \tfrac{T}{2} + \varepsilon \leq t \leq T $. Let $ v(t) = \bar{\Phi}(t,\varepsilon) \alpha + {v}_2(t) $.
	Notice that $ v(t) $ fulfills \eqref{eqProofTwoNeedlesVarEq2} such that 
	\begin{align}
		x(T) &= x^*(T) + \varepsilon \bar{\Phi}(T,\tfrac{T}{2}+\varepsilon) v(\tfrac{T}{2}+\varepsilon) + \mathcal{O}(\varepsilon^2) \label{eqProofTwoNeedlesAtT1} \\
		     &= x^*(T) + \varepsilon \bar{\Phi}(T,\tfrac{T}{2}+\varepsilon) \big( \bar{\Phi}(\tfrac{T}{2}+\varepsilon,\varepsilon) \alpha - \alpha \big)  + \mathcal{O}(\varepsilon^2).  \nonumber
	\end{align}
	We will now investigate how $ \bar{\Phi}(t,t_0) $ and $ \Phi(t,t_0) $ are related.
	Let $ \Phi_2(t,t_0) $ denote the state-transition matrix corresponding to \eqref{eqProofTwoNeedlesVarEq2}
	when $ u_1(t) = -1 $ and notice that $ \Phi(t,t_0) $ is the state-transition matrix for the
	case of $ u_1(t) = 1 $. Then we have
% 	\begin{align}
% 		\bar{\Phi}(t,t_0) &= \begin{cases}
% 							\Phi(t,t_0)                                   & \text{if } 0 \leq t < \tfrac{T}{2}, 0 \leq t_0 \leq \tfrac{T}{2} \\
% 							\Phi(t,\tfrac{T}{2}) \Phi_2(\tfrac{T}{2},t_0) & \text{if } 0 \leq t \leq \tfrac{T}{2}, \tfrac{T}{2} \leq t_0 \leq T \\
% 							\Phi_2(t,\tfrac{T}{2}) \Phi(\tfrac{T}{2},t_0) & \text{if } \tfrac{T}{2} \leq t \leq T, 0 \leq t_0 < \tfrac{T}{2} \\
% 							\Phi_2(t,t_0)                                 & \text{if } \tfrac{T}{2} \leq t \leq T, \tfrac{T}{2} \leq t_0 \leq T .
% 		                     \end{cases} \label{eqDefPhiBar}
% 	\end{align}
	\begin{flalign}
		\bar{\Phi}(t,t_0) &= \begin{cases}
							\Phi(t,t_0)                                   & \hspace*{-0.6em} \text{if } t \in [0,\tfrac{T}{2}), t_0 \in [0,\tfrac{T}{2}] \\ %      0 \leq t < \tfrac{T}{2}, 0 \leq t_0 \leq \tfrac{T}{2} \\
							\Phi(t,\tfrac{T}{2}) \Phi_2(\tfrac{T}{2},t_0) & \hspace*{-0.6em} \text{if } t \in [0,\tfrac{T}{2}], t_0 \in [\tfrac{T}{2},T] \\ %  0 \leq t \leq \tfrac{T}{2}, \tfrac{T}{2} \leq t_0 \leq T \\
							\Phi_2(t,\tfrac{T}{2}) \Phi(\tfrac{T}{2},t_0) & \hspace*{-0.6em} \text{if } t \in [\tfrac{T}{2},T], t_0 \in [0,\tfrac{T}{2}) \\ %  \tfrac{T}{2} \leq t \leq T, 0 \leq t_0 < \tfrac{T}{2} \\
							\Phi_2(t,t_0)                                 & \hspace*{-0.6em} \text{if } t \in [\tfrac{T}{2},T], t_0 \in [\tfrac{T}{2},T]    %  \tfrac{T}{2} \leq t \leq T, \tfrac{T}{2} \leq t_0 \leq T .
		                     \end{cases} \hspace*{-10pt} & \label{eqDefPhiBar}
	\end{flalign}
	Now, since \eqref{eqProofTwoNeedlesVarEq2} is a scalar linear time varying
	differential equation, we have 
	$
		\Phi(t,t_0) = \exp( \int_{t_0}^{t} \tfrac{\partial F}{\partial x}\big( x^*(\tau) \big) d\tau .
	$
	Similarly
	\begin{align}
		& \Phi_2(t,t_0) \nonumber \\
				   &= \exp( - \int_{t_0}^{t} \tfrac{\partial F}{\partial x}\big( x^*(\tau) \big) d\tau 
		              = \exp( - \int_{t_0}^{t} \tfrac{\partial F}{\partial x}\big( x^*(T-\tau) \big) d\tau \nonumber \\
		              &= \exp( \int_{T-t_0}^{T-t} \tfrac{\partial F}{\partial x}\big( x^*(s) \big) ds 
		              = \Phi(T-t,T-t_0) \label{eqPhi2}
	\end{align}
	where we used that $ x^*(t) = x^*(T-t) $ since $ x^*(t) $ is just going
	back and forth along $ F(x) $. Using \eqref{eqDefPhiBar} and \eqref{eqPhi2}
	in \eqref{eqProofTwoNeedlesAtT1} we obtain with $ x^*(T) = x_0 $
	\begin{align}
		x(t) &= x_0 + \varepsilon \alpha \big( \Phi_2(T,\tfrac{T}{2}) \Phi(\tfrac{T}{2},\varepsilon) - \Phi_2(T,\tfrac{T}{2}+\varepsilon) \big) + \mathcal{O}(\varepsilon^2) \nonumber \\
		     &= x_0 + \varepsilon \alpha \big( \Phi(0,\varepsilon) - \Phi(0, \tfrac{T}{2}-\varepsilon) \big) +  \mathcal{O}(\varepsilon^2) \nonumber \\
		     &= x_0 + \varepsilon \alpha \Phi(0,\varepsilon) \big( 1 - \Phi(\varepsilon, \tfrac{T}{2}-\varepsilon) \big) +  \mathcal{O}(\varepsilon^2) \nonumber \\
		     &= x_0 + \varepsilon \alpha \Phi(0,\varepsilon) \int_{\varepsilon}^{\tfrac{T}{2}-\varepsilon} \tfrac{\partial}{\partial \tau} \Phi(\tau, \tfrac{T}{2}-\varepsilon) d\tau +  \mathcal{O}(\varepsilon^2) \nonumber \\
		     &= x_0 + \varepsilon \alpha \Phi(0,\varepsilon) \int_{\varepsilon}^{\tfrac{T}{2}-\varepsilon} \tfrac{\partial F}{\partial x} \big( x^*(\tau) \big) \Phi(\tau, \tfrac{T}{2}-\varepsilon) d\tau \nonumber \\
		     &\phantom{=}+  \mathcal{O}(\varepsilon^2) 
	\end{align}
	where in the last step we used the differentiation property of the 
	state-transition matrix, see e.g \cite{kailath1980linear}.	
\end{proof}
\begin{remark}\label{remarkExtensionTwoNeedles}
	Equation \eqref{eqApproximationTwoNeedles1} gives an approximation at $ t = T $.
	However, due to the periodicity, it can as well be extended to arbitrary integer multiples of $ T $
	which leads to
% 	\begin{align}
% 		x\big((k+1)T\big) = x(kT) \nonumber \\ 
% 		+ \varepsilon \alpha \Phi(kT,kT+\varepsilon) \int_{kT+\varepsilon}^{kT+\tfrac{T}{2}-\varepsilon} \frac{\partial F}{\partial x} \big( x^*(\tau) \big) \Phi(\tau,kT+\tfrac{T}{2}-\varepsilon) \, d\tau \nonumber \\
% 			\phantom{=}+ \mathcal{O}(\varepsilon^2) \label{eqTwoNeedlesIteration}
% 	\end{align}
	\begin{align}
		&x\big((k+1)T\big) =  \label{eqTwoNeedlesIteration} \\ 
		& x(kT) 
		 + \varepsilon \alpha \Phi(kT,t_{k1}) \int_{t_{k1}}^{t_{k2}} \tfrac{\partial F}{\partial x} \big( x^*(\tau) \big) \Phi(\tau,t_{k2}) \, d\tau 
			+ \mathcal{O}(\varepsilon^2) \nonumber
	\end{align}
	where $ k \in \mathbb{N} $, $ t_{k1} := kT+\varepsilon $, $ t_{k2} := kT+\tfrac{T}{2}-\varepsilon $.
\end{remark}
% \begin{remark}
% 	The inputs $ u_1, u_2 $ as defined in \eqref{eqInputSequence1_u1} and \eqref{eqInputSequence1_u2}
% 	can be smoothly approximated by  $ {u}_1 = \big( {\sin(\tfrac{2\pi}{T}t)} \big)^{\tfrac{1}{2N+1}} $ and
% 	$ u_2 = \big( \cos(\tfrac{2\pi}{T}t) \big)^{2N+1} $ for $ N \in \mathbb{N} $ sufficiently large.
% \end{remark}
Eq. \eqref{eqApproximationTwoNeedles1} gives insight into the gradient approximation process and
shows that system \eqref{eqNonlinearInputAffineSys} with inputs \eqref{eqInputSequence1_u1}
and \eqref{eqInputSequence1_u2} approximately moves along a weighted
average of the gradient of $ F $. Thus, depending on the sign of $ \alpha $, this system does
a modified gradient descent or ascent. This is also comparable to what is done
in stochastic approximation (\cite{spall1992multivariate}) where one uses the sum of
several gradients. \\
Notice that the approximation \eqref{eqApproximationTwoNeedles1} is valid for
small $ \varepsilon $ but $ T $ does not have to be small which is in contrast
to existing results. In the following
we consider the case where $ T $ is of the same order of magnitude as 
$ \varepsilon $. In particular, we consider $ T = 8\varepsilon $.
\begin{snugshade}
\begin{lemma}\label{lemmaDeltaEqualEpsilon}
	Consider \eqref{eqNonlinearInputAffineSys} together with the input sequence
	as defined by \eqref{eqInputSequence1_u1} and \eqref{eqInputSequence1_u2}.
	Let $ x^*(t) $ denote the solution of \eqref{eqNonlinearInputAffineSys} when
	$ u_1(t) $ as defined by \eqref{eqInputSequence1_u1} and $ u_2(t) \equiv 0 $.
	Suppose $ T = 8\varepsilon $. Then
	\begin{align}
		x(T) &= x_0 + \varepsilon^2 \alpha \bigg( \tfrac{\partial F}{\partial x}\big(x^*(\tfrac{T}{2})\big)  +   \tfrac{\partial F}{\partial x}\big(x_0\big)  \bigg) + \mathcal{O}(\varepsilon^3).
		\label{eqApproxDeltaEqualEpsilon}
	\end{align}
\end{lemma}
\end{snugshade}
\begin{proof}
	In case of $ T = 8\varepsilon $ we have from \eqref{eqApproximationTwoNeedles1}
	\begin{align}
		x(T) &= x_0 + \varepsilon \alpha \Phi(0,\varepsilon) \int\limits_{\varepsilon}^{3\varepsilon} \tfrac{\partial F}{\partial x} \big( x^*(\tau) \big) \Phi(\tau,3\varepsilon) \, d\tau + \mathcal{O}(\varepsilon^2) . \label{eqProofLemmaTwoNeedles1}
	\end{align}
	Expanding the integrand into a Taylor series about $ \tau = 4\varepsilon $ we obtain
	\begin{align}
		 &\int_{\varepsilon}^{3\varepsilon} \tfrac{\partial F}{\partial x} \big( x^*(\tau) \big) \Phi(\tau,3\varepsilon) \, d\tau \nonumber \\
		=\, & \int_{\varepsilon}^{3\varepsilon} \tfrac{\partial F}{\partial x} \big( x^*(4\varepsilon) \big) \Phi(4\varepsilon,3\varepsilon) + \mathcal{O}(\tau-4\varepsilon) \, d\tau \nonumber \\
		=\, & 2 \varepsilon \tfrac{\partial F}{\partial x} \big( x^*(4\varepsilon) \big) \Phi(4\varepsilon,3\varepsilon) + \mathcal{O}(\varepsilon^2) . \label{eqProofLemmaTwoNeedles2}
	\end{align}
	Moreover, by a Taylor expansion of $ \Phi(4\varepsilon,t_0) $ about \mbox{$ t_0=4\varepsilon$}, it is
	\begin{align}
		\Phi(4\varepsilon,3\varepsilon) = \Phi(4\varepsilon,4\varepsilon) - \tfrac{d}{dt_0} \Phi(4\varepsilon,t_0)_{| t_0=4\varepsilon} \varepsilon + \mathcal{O}(\varepsilon^2). \label{eqProofLemmaTwoNeedles3}
	\end{align}
	Equivalently, we have
	\begin{align}
		\Phi(0,\varepsilon) = \Phi(0,0) + \tfrac{d}{dt_0} \Phi(0,t_0)_{| t_0=0} \varepsilon + \mathcal{O}(\varepsilon^2). \label{eqProofLemmaTwoNeedles4}
	\end{align}
	Hence we obtain using \eqref{eqProofLemmaTwoNeedles2} - \eqref{eqProofLemmaTwoNeedles4}
	in \eqref{eqProofLemmaTwoNeedles1}
	\begin{align}
		x(T) &= x_0 + 2 \varepsilon^2 \alpha \tfrac{\partial F}{\partial x} \big( x^*(4\varepsilon) \big) + \mathcal{O}(\varepsilon^3).
	\end{align}
	By \eqref{eqProofTwoNeedlesBeforeSecondNeedle} it is $ x^*(4\varepsilon) = x(\tfrac{T}{2}) + \mathcal{O}(\varepsilon) $
	but also $ x^*(4\varepsilon) = x_0 + \mathcal{O}(\varepsilon) $. Thus
	\begin{align}
		x(T) &= x_0 + 2 \varepsilon^2 \alpha \tfrac{\partial F}{\partial x} \big( x(\tfrac{T}{2}) \big) + \mathcal{O}(\varepsilon^3) \\
		     &= x_0 + 2 \varepsilon^2 \alpha \tfrac{\partial F}{\partial x} \big( x_0 \big) + \mathcal{O}(\varepsilon^3)
	\end{align}
	such that -- by adding both expressions -- we finally arrive at \eqref{eqApproxDeltaEqualEpsilon}.
% 	\begin{align}
% 		x(T) &= x_0 + \varepsilon^2 \alpha \bigg( \tfrac{\partial F}{\partial x}\big(x(\tfrac{T}{2})\big)  +   \tfrac{\partial F}{\partial x}\big(x_0\big)  \bigg) + \mathcal{O}(\varepsilon^3).
% 	\end{align}
\end{proof}
\begin{remark}
	\Cref{lemmaDeltaEqualEpsilon} is consistent with the well-known fact that an
	input-affine system
	$
		\dot{x} = g_1(x) u_1(t) + g_2(x) u_2(t)
	$
	with $u_1$, $u_2$ as defined by \eqref{eqInputSequence1_u1} and \eqref{eqInputSequence1_u2} 
	where $ T = 8 \varepsilon $ approximately moves into the direction of the Lie bracket
	$ [ g_1, g_2 ](x) $, see e.g. \cite{brockett1976} or \cite{duerr2013LieBracket}
	in terms of extremum seeking. Here, it is $ g_1(x) = F(x) $
	and $ g_2(x) = 1 $ such that we obtain for the Lie bracket
	\begin{align}
		[g_1,g_2](x) = \tfrac{\partial g_2}{\partial x}(x) g_1(x) - \tfrac{\partial g_1}{\partial x}(x) g_2(x)  
		             = - \tfrac{\partial F}{\partial x} (x) .
	\end{align}
\end{remark}
\begin{remark}
	In case of $ T = 4\varepsilon $ the integral in \eqref{eqApproximationTwoNeedles1} vanishes
	such that there are no first order terms. 	Thus, the ratio between $ \varepsilon $ 
	and $ T $ is important but nevertheless $T$ does not necessarily have to be $8\varepsilon$.
\end{remark}
In comparison to common approximation formulas our result \eqref{eqApproxDeltaEqualEpsilon}
not only includes the gradient of $ F $ evaluated at the start point but also
at the point where $ u_1 $ changes its sign, i.e. at
the two points where $ x^* $ reverses its direction.
Usually, this term disappears by including it in the higher order terms.

%%%%%%%%%%%%%%%%%%%%%%%%%%%%%%%%%%%%%%%%%%%%%%%%%%%%%%%%%%%%%%%%%%%%%%%%%%%%%%%
\subsection{Dither signals composed of infinitely many needles}\label{subsectionManyNeedles}
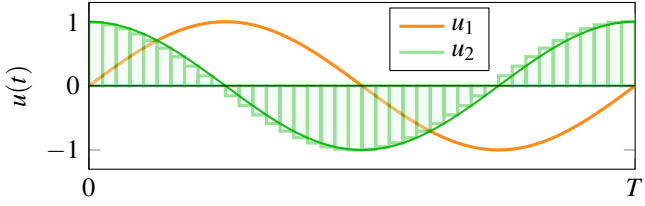
\begin{figure}
	\vspace*{0.2cm}
	%\def\N{40}						% number of needles
%\FPeval{\eps}{2*pi/{\N}}			% width of one needle
%\def\omegaVal{1}				% frequency of sinusoidal signal
\colorlet{u1Color}{orange}			% color for u1
\begin{tikzpicture}[>=latex]
	\begin{axis}[
			width=\columnwidth,
			height=3.8cm,
			xmin=0, 
			xmax=6.2832,
			ymin=-1.3, 
			ymax=1.3,
			disabledatascaling,
			xlabel=$ $,
			ylabel=$u(t)$,
			xtick={0,6.28},
			xticklabels={ $0$,$T$  },
			ytick={-1,0,1},
			yticklabels={$-1$,$0$,$1$,},
			ylabel style={yshift=-0.3cm},
			legend style={at={(0.55,0.97)},anchor=north west,nodes=right},
			grid=none]
		% plot u1
		\addplot[very thick,color=u1Color,opacity=0.9,domain=0:6.2832,samples=100] { sin(1*deg(x)) }; 
		\addlegendentry{$u_1$};
		% plot u2: approximation and smooth version
		\addplot[very thick,color=green!70!black,opacity=0.4,fill=green!70!black!10,domain=0:6.2832,samples=41, ybar interval] { cos(1*deg(x+0.1571)) };
		\addlegendentry{$u_2$};
		\addplot[thick,color=green!70!black,opacity=0.9,domain=0:6.2832,samples=100,forget plot] { cos(1*deg(x)) };
		% coordinate axis
		\draw[-] (axis cs: -1,0) -- (axis cs: 6.28,0);
		\draw[-] (axis cs: 0,-1.5) -- (axis cs: 0,1.5);
	\end{axis} 
\end{tikzpicture}\vspace*{-1em}
	\caption{Illustration of the approximation of the input function $u_1$ by many needles as used in \Cref{theoremManyNeedles}.}
	\label{figInputSequence2}
\end{figure}

Up to now we have considered the very special input sequence given by \eqref{eqInputSequence1_u1}
and \eqref{eqInputSequence1_u2}. In the following we consider the same nonlinear input-affine system \eqref{eqNonlinearInputAffineSys}
with more general inputs $ u_1 $ and $ u_2 $. More precisely, we impose the following assumptions
on the input functions $ u_1, u_2 : \mathbb{R} \to \mathbb{R} $:
\begin{itemize}
	\item[\textbf{A1}] The functions $ u_1, u_2 $ are piecewise continuous and bounded.	\label{assA1}
	\item[\textbf{A2}] The functions $ u_1, u_2 $ are $T$-periodic and have zero mean.	\label{assA2}
\end{itemize}\vspace*{-0.5em}
For example, this includes the well-known case of trigonometric input functions, i.e.
$ u_1(t) = \sqrt{\omega} \cos(\omega t) $ and \mbox{$ u_2(t) = \sqrt{\omega} \sin(\omega t) $}
that was considered e.g. in \cite{kurzweil1987limit} or more recently in \cite{duerr2013LieBracket}
in the context of extremum seeking. 
Loosely spoken, the idea is to approximate the input $ u_2 $ by infinitely many
needles to go along the same lines as in the previous case, see \Cref{figInputSequence2}.
\myComment{How can the following Theorem be interpreted and how exactly is it
related to the representation with iterated integrals?}
\begin{snugshade}
\begin{theorem}\label{theoremManyNeedles}
	Suppose that Assumptions \hyperref[assA1]{A1} and \hyperref[assA2]{A2} hold.
	Consider \eqref{eqNonlinearInputAffineSys} together with $ u_2(t) $ defined by
	\begin{align*}
		u_2(t) = u_2( t_{i+1} ) \qquad \text{for } t \in [t_i, t_{i+1}), t_{i+1} \leq T 
	\end{align*}
	where $ t_i = \varepsilon i $, $ \varepsilon = \tfrac{T}{N} $, $N \in \mathbb{N} $, $ i = 0,1,\dots,N-1$.
	Let $ x^*(t) $ denote the solution of 
	\eqref{eqNonlinearInputAffineSys} for $u_2(t) \equiv 0 $ and $u_1(t)$
	as defined by Assumptions \hyperref[assA1]{A1} and \hyperref[assA2]{A2}
	and suppose that $ x^*(t) $ exists on $ [0,T] $. 
	Let further $ \Phi(t,t_0) $ denote the state-transition matrix
	corresponding to
	\begin{align}
		\dot{v}_1(t) = \tfrac{\partial F}{\partial x}\big( x^*(t) \big) u_1(t) v_1(t) \label{eqManyNeedlesVariationalEquation}
	\end{align}
	with initial time $ t_0 $. Then, if $N$ is finite,
% 	\begin{align}
% 		&x(T) = x^*(T) \label{eqTheoremManyNeedlesNoLimit} \\
% 		&+ \varepsilon \sum\limits_{i=0}^{N-1} \bigg( \int\limits_{t_{i+1}}^{T} \tfrac{\partial F}{\partial x} \big(x^*(\tau)\big) u_1(\tau) \Phi\big( \tau,t_{i+1}\big) d\tau + 1 \bigg) u_2\big( t_{i+1} \big) + \mathcal{O}(\varepsilon^2). \nonumber
% 	\end{align}
% 	\begin{align}
% 		&x(T)= \nonumber \\
% 		& x^*(T) + \varepsilon \sum\limits_{i=0}^{N-1} \big( \int\limits_{t_{i+1}}^{T} \tfrac{\partial F}{\partial x} \big(x^*(\tau)\big) u_1(\tau) \Phi\big( \tau,t_{i+1}\big) d\tau + 1 \big) u_2\big( t_{i+1} \big) \nonumber \\
% 		&+ \mathcal{O}(\varepsilon^2). \label{eqTheoremManyNeedlesNoLimit}
% 	\end{align}
	\begin{align}
		&x(T)= x^*(T) + \mathcal{O}(\varepsilon^2)  \label{eqTheoremManyNeedlesNoLimit} \\
		& + \varepsilon \sum\limits_{i=0}^{N-1} \big( \int\limits_{t_{i+1}}^{T} \tfrac{\partial F}{\partial x} \big(x^*(\tau)\big) u_1(\tau) \Phi\big( \tau,t_{i+1}\big) d\tau + 1 \big) u_2\big( t_{i+1} \big) 
		. \nonumber
	\end{align}
	Moreover,
	\begin{align}
		\lim\limits_{N\to\infty} x(T) &= x^*(T)  % + \int\limits_0^T u_2(t) dt \\
		                                       + \int\limits_{0}^{T} \int\limits_{t}^{T}  \tfrac{\partial F}{\partial x} \big( x^*(\tau) \big) u_1(\tau) \Phi( \tau, t ) u_2( t ) \, d\tau \, dt.\label{eqTheoremManyNeedlesLimit}  
	\end{align}
\end{theorem}
\end{snugshade}
\begin{proof}
	We use again the results presented in \Cref{subsectionPrelimVariationalEquation}. With \eqref{eqPrelimAfterNeedle}
	we have at the end of the first needle  
	$
		x(\varepsilon) = x^*(\varepsilon) + \varepsilon u_2(\varepsilon) + \mathcal{O}(\varepsilon^2).
	$
	Let $ \bar{x}(t) $ denote the solution in case we would not apply the second needle
	at $ t = \varepsilon $. Then
	\begin{align}
		\bar{x}(2\varepsilon) &= x^*(2\varepsilon) + \varepsilon v_1(2\varepsilon) + \mathcal{O}(\varepsilon^2) \nonumber \\
		                      &= x^*(2\varepsilon) + \varepsilon \Phi(2\varepsilon,\varepsilon) u_2(\varepsilon) + \mathcal{O}(\varepsilon^2)
	\end{align}
	and hence when the second needle is applied we obtain using again \eqref{eqPrelimAfterNeedle}
	\begin{flalign}
		x(2\varepsilon) &= \bar{x}(2\varepsilon) + \varepsilon u_2(2\varepsilon) + \mathcal{O}(\varepsilon^2) \nonumber \\
		                &= x^*(2\varepsilon) + \varepsilon \Phi(2\varepsilon,\varepsilon) u_2(\varepsilon) + \varepsilon u_2(2 \varepsilon) + \mathcal{O}(\varepsilon^2). \hspace*{-3em} &
	\end{flalign}
	Here, the first term involving the state-transition matrix is the propagation of the
	first needle whereas the other term linear in $ \varepsilon $ is the second needle.
	Going on along these lines we obtain
	\begin{align}
		x(3\varepsilon) &= x^*(3\varepsilon) + \Phi(3\varepsilon,2\varepsilon)\big( \Phi(2\varepsilon,\varepsilon)\varepsilon u_2(\varepsilon) 
															+ \varepsilon u_2(2\varepsilon) \big) \nonumber \\
					&\phantom{=}		+ \varepsilon u_1(3\varepsilon) + \mathcal{O}(\varepsilon^2) \nonumber \\
				     &= x^*(3\varepsilon) + \varepsilon \Phi(3\varepsilon,\varepsilon) u_2(\varepsilon) + \varepsilon \Phi(3\varepsilon,2\varepsilon) u_2(2\varepsilon) \nonumber \\
				     &\phantom{=} + \varepsilon u_2(3\varepsilon) + \mathcal{O}(\varepsilon^2).
	\end{align}
	Generalizing this we have
	\begin{align}
		x(k\varepsilon) &= x^*(k\varepsilon) + \varepsilon \sum\limits_{i=0}^{k-1} \Phi(k\varepsilon,t_{i+1}) u_2( t_{i+1} ) + \mathcal{O}(\varepsilon^2)
	\end{align}
	where $ k \in \mathbb{N} $ and $ t_i = \varepsilon i $. Thus, we have after one period, i.e. at $ T = N\varepsilon $,
	\begin{align}
		x(T) = x^*(T) + \varepsilon \sum\limits_{i=0}^{N-1} \Phi(T,t_{i+1}) u_2( t_{i+1} ) + \mathcal{O}(\varepsilon^2) .  \label{eqProofManyNeedlesAtT}
	\end{align}
	We will now take a closer look at the sum. It is
	\begin{align}
		& \varepsilon \sum\limits_{i=0}^{N-1} \Phi(T,t_{i+1}) u_2( t_{i+1} ) \\
% 		=& \sum\limits_{i=0}^{N-1} \bigg( \Phi\big(T,(i+1)\varepsilon\big) - \Phi\big( (i+1)\varepsilon, (i+1)\varepsilon \big) + \Phi\big( (i+1)\varepsilon, (i+1)\varepsilon \big)   \bigg) \varepsilon u_1\big( (i+1)\varepsilon \big) \\
		=& \varepsilon \sum\limits_{i=0}^{N-1} \bigg( \int\limits_{t_{i+1}}^{T} \tfrac{\partial\Phi}{\partial\tau} (\tau,t_{i+1}) d\tau + 1 \bigg) u_2( t_{i+1} ) \nonumber \\
		=& \varepsilon \sum\limits_{i=0}^{N-1} \bigg( \int\limits_{t_{i+1}}^{T} \tfrac{\partial F}{\partial x} \big(x^*(\tau)\big) u_1(\tau) \Phi\big( \tau,t_{i+1}) d\tau + 1 \bigg) u_2( t_{i+1} ) \nonumber
	\end{align}
	where we used in the last step that the state-transition matrix $\Phi$ fulfills
	the variational equation \eqref{eqManyNeedlesVariationalEquation}. This, together
	with \eqref{eqProofManyNeedlesAtT}, proves 	\eqref{eqTheoremManyNeedlesNoLimit}.
	Next we look at the case when $ N $ tends to infinity.
	We consider first the part without he integral, i.e.
	\begin{align}
		\lim\limits_{N \to \infty} \sum\limits_{i=0}^{N-1} \varepsilon u_2( t_{i+1} ) &= \lim\limits_{N \to \infty} \sum\limits_{i=0}^{N-1} ( t_{i+1} - t_i ) u_2( t_{i+1} ).
	\end{align}
     This is the limit of a Riemann sum (see e.g \cite[Chapter 3]{trench2003introduction}) with partition $ \lbrace t_i \rbrace $ 
     such that
	\begin{align}
		\lim\limits_{N \to \infty} \sum\limits_{i=0}^{N-1} \varepsilon u_2( t_{i+1} ) &= \int\limits_0^T u_2(t) dt .
	\end{align}
	Since $ u_2 $ has zero mean by Assumption \hyperref[assA2]{A2}, this integral vanishes. 
	For the second part of the sum we calculate similarly
	\begin{align}
		&\lim\limits_{N\to\infty} \sum\limits_{i=0}^{N-1} \int\limits_{t_{i+1}}^{T} \tfrac{\partial F}{\partial x} \big( x^*(\tau) \big) u_1(\tau) \Phi\big( \tau, t_{i+1} \big) \varepsilon u_2( t_{i+1} ) d\tau \nonumber \\
		=& \lim\limits_{N\to\infty} \sum\limits_{i=0}^{N-1} \int\limits_{t_{i+1}}^{T} \tfrac{\partial F}{\partial x} \big( x^*(\tau) \big) u_1(\tau) \Phi\big( \tau, t_{i+1} \big) ( t_{i+1} - t_i ) u_2\big( t_{i+1} \big) d\tau \nonumber \\
		=& \int\limits_{0}^{T} \int\limits_{t}^{T}  \tfrac{\partial F}{\partial x} \big( x^*(\tau) \big) u_1(\tau) \Phi( \tau, t ) u_2( t ) d\tau \, dt
	\end{align}
	where we once more used that we have a Riemann sum here that converges to the Riemannian
	integral.
\end{proof}
\begin{remark}
	\Cref{theoremManyNeedles} also applies with a slight modification if Assumption \hyperref[assA2]{A2} is not fulfilled. 
	However, periodicity is required to extend the approximation as explained in \Cref{remarkExtensionTwoNeedles}.
	Further, the mean of $ u_2(t) $ appears in \eqref{eqTheoremManyNeedlesLimit} if it does not vanish.
\end{remark}
\begin{remark}\label{remarkTrigonometricFunctions}
	For the standard case of trigonometric functions $ u_1(t) = \sqrt{\omega} \cos(\omega t) $
	and $ u_2(t) = \sqrt{\omega} \sin(\omega t) $, $ \omega = \tfrac{2\pi}{T} $, one can verify 
	using \eqref{eqTheoremManyNeedlesLimit} the result as presented in \cite{duerr2013LieBracket} 
	where the case that $ T $ tends to zero is considered. However, due to space limitations,
	we skip the calculations here.
	Notice also that the related functions $ {u}_1(t) = \big( {\sin(\tfrac{2\pi}{T}t)} \big)^{\tfrac{1}{2N+1}} $ and
 	$ u_2(t) = \big( \cos(\tfrac{2\pi}{T}t) \big)^{2N+1} $ are a smooth approximation of
 	$ u_1, u_2 $ as defined by \eqref{eqInputSequence1_u1} and \eqref{eqInputSequence1_u2}
 	for $ N \in \mathbb{N} $ sufficiently large.
	% \begin{remark}
% 	The inputs $ u_1, u_2 $ as defined in \eqref{eqInputSequence1_u1} and \eqref{eqInputSequence1_u2}
% 	can be smoothly approximated by  $ {u}_1 = \big( {\sin(\tfrac{2\pi}{T}t)} \big)^{\tfrac{1}{2N+1}} $ and
% 	$ u_2 = \big( \cos(\tfrac{2\pi}{T}t) \big)^{2N+1} $ for $ N \in \mathbb{N} $ sufficiently large.
% \end{remark}
\end{remark}
\begin{remark}
	\Cref{theoremTwoNeedles} is a special case of \Cref{theoremManyNeedles}. 
% 	For input functions
% 	$u_1, u_2 $ as defined by \eqref{eqInputSequence1_u1} and \eqref{eqInputSequence1_u2} the
% 	approximation \eqref{eqTheoremManyNeedlesNoLimit} reduces to \eqref{eqApproximationTwoNeedles1}.
	This can be seen by evaluating \eqref{eqTheoremManyNeedlesNoLimit} for $u_1, u_2$ as given by
	\eqref{eqInputSequence1_u1} and \eqref{eqInputSequence1_u2} and using the differentiation
	property of the state-transition matrix $ {\Phi} $ to get rid of the integral. Notice that 
	$ {\Phi} $ plays the role of $ \bar{\Phi} $ as defined in the proof of \Cref{theoremManyNeedles}. By that, 
	one ends up with \eqref{eqProofTwoNeedlesAtT1} and following the rest of the proof of
	\Cref{theoremTwoNeedles} the relation becomes clear.
\end{remark}

%%%%%%%%%%%%%%%%%%%%%%%%%%%%%%%%%%%%%%%%%%%%%%%%%%%%%%%%%%%%%%%%%%%%%%%%%%%%%%%
\subsection{An accelerated gradient algorithm}\label{subsectionNewAlgorithm}
% \begin{figure*}[ht]
% %	\includegraphics[width=\textwidth]{figures/standaloneComparisonComplete.pdf}
% 	\def\figurewidth{0.44\textwidth}
% 	\def\figureheight{0.22\textwidth}
% 	\input{figures/comparisonCompleteNewReduced.tex}
% 	\caption{A comparison between the original system (green), i.e. system \eqref{eqNonlinearInputAffineSys} with inputs \eqref{eqInputSequence1_u1} and \eqref{eqInputSequence1_u2}, the approximation \eqref{eqApproximationTwoNeedles1} (orange)
%  	         and a gradient descent algorithm (blue) in the case of quadratic $ F(x) $ ($b=2$,$c=3$,$\varepsilon=0.00001$,$T=1.3$,$x_{min}=-1$,$\alpha=-10$).
%  	         On the left hand side the thick green line depicts the original system evaluated at multiples of the period length;
%  	         the shaded area shows the original system jumping up and down very fast. A zoomed plot that illustrates this jumping
%  	         can be seen on the right hand side.}\vspace*{-1.4em}
%  	\label{figSimulationResultsQuadraticCase}
% \end{figure*}
In the following we present a new continuous-time optimization algorithm. 
The algorithm is presented in the following Theorem and is
motivated by \Cref{lemmaDeltaEqualEpsilon} as
we will explain in \Cref{remarkNewAlgorithmMotivation}. 
%%%%%%%%%%%%%%%%%%%%%%%%%%%%%%%%%%%%%%%%%%%%%%%%%%%%%%%%%%%%%%%%%%%%%%%%%%%%%%%%%%%%%%%%%%%%%%%%%%%%%
\begin{snugshade*}
\begin{theorem}\label{theoremNewAlgorithm}
	Consider 
	\begin{align}
		\begin{split}
			\dot{z}_1 &=  z_2 \\
			\dot{z}_2 &= -k z_2 - c_1 \nabla F(z_1) - c_2 \nabla F(z_1 + \gamma z_2)
		\end{split} \label{eqNewAlgorithm}
	\end{align}
	where $ k,c_1,c_2,\gamma > 0 $ and $ z_1, z_2 \in \mathbb{R}^n $. Suppose
	$ F: \mathbb{R}^n \to \mathbb{R} $, $ F \in \mathcal{C}^1 $ is locally convex on an open convex set
	$ \mathcal{D} \subseteq \mathbb{R}^n $ and has a unique isolated minimum on $ \mathcal{D} $
	at $ z^* \in \mathcal{D} $. Then the equilibrium $ \begin{bmatrix} {z^*}^\top & 0 \end{bmatrix}^\top $
	of \eqref{eqNewAlgorithm} is locally asymptotically stable. 
	Moreover, if $ \mathcal{D} = \mathbb{R}^n $ and $F$ is radially unbounded, then the equilibrium is 
	globally asymptotically stable for \eqref{eqNewAlgorithm}.
\end{theorem}
\end{snugshade*}
\begin{proof}
	It is straight forward to verify that $ \bar{z}_1 = z^*,\bar{z}_2 = 0 $
	is an equilibrium of \eqref{eqNewAlgorithm} since $ \nabla F(\bar{z}_1) = \nabla F(\bar{z}_1+\gamma \bar{z}_2) = 0 $.
	Consider now the Lyapunov function candidate
	\begin{align}
		V = \tfrac{1}{2} z_2^\top z_2 + (c_1+c_2) F(z_1) - (c_1+c_2) F(z^*)
	\end{align}
	which is positive definite for $ z_1 \in \mathcal{D} $, $ z_2 \in \mathbb{R}^n $
	and attains its minimum at the considered equilibrium.
	The derivative of $ V $ along the trajectories of \eqref{eqNewAlgorithm} is
	given by
	\begin{align}
		\dot{V} &= z_2^\top \dot{z}_2 + (c_1+c_2) \dot{z}_1^\top \nabla F(z_1) \nonumber \\
	             &= z_2^\top \big( -k z_2 - c_1 \nabla F(z_1) - c_2 \nabla F(z_1 + \gamma z_2) \big) \nonumber \\
	             &\phantom{=} + (c_1+c_2) z_2^\top \nabla F(z_1) \nonumber \\
	             &= -k \Vert z_2 \Vert_2^2 - c_2 z_2^\top \big( \nabla F(z_1 + \gamma z_2) - \nabla F(z_1) \big).
	\end{align}
	Since $ F $ is assumed to be locally convex, we have that
	\begin{align*}
		(\alpha-\beta)^\top \big( \nabla F(\alpha) - \nabla F(\beta) \big) \geq 0 \qquad  \text{for all } \alpha, \beta \in \mathcal{D}.
	\end{align*}
	Here, $ \alpha = z_1 + \gamma z_2 $ and $ \beta = z_1 $ such that $ \alpha-\beta = \gamma z_2 $ and hence,
	since $ \gamma > 0 $, we have
	$
		z_2^\top \big( \nabla F(z_1 + z_2) - \nabla F(z_1) \big) \geq 0.
	$
	Thus, $ \dot{V} \leq 0 $ for $ z_1 \in \mathcal{D} $, $z_2 \in \mathbb{R}^n $. In particular,
	$ \dot{V} = 0 $ for $ z_2 = 0 $. However, putting $ z_2 \equiv 0 $ into
	\eqref{eqNewAlgorithm}, it follows that $ z_1 \equiv z^* $ such that by the invariance principle of
	Krasovskii-LaSalle we conclude that $ (z^*,0) $ is locally asymptotically stable for \eqref{eqNewAlgorithm}.
	Equally, in case $ \mathcal{D} = \mathbb{R}^n $, $ V $ is radially unbounded and $ \dot{V} \leq 0 $
	for all $ z_1,z_2 \in \mathbb{R}^n $ such that -- again by the invariance principle -- 
	the equilibrium is globally asymptotically stable.
\end{proof}
\begin{remark}\label{remarkNewAlgorithmMotivation}
	The algorithm is motivated by the result of \Cref{lemmaDeltaEqualEpsilon} as we will
	show in the following. 
	%It should be mentioned that this is not a derivation but rather 
	%an explanation of how we came up with the algorithm. 
	Consider again \eqref{eqApproxDeltaEqualEpsilon}
	and let $ x^{(k)} := x(k\tfrac{T}{2}) $, $ k \in \mathbb{N} $.
	Then we have as a generalization of \eqref{eqApproxDeltaEqualEpsilon} 
	\begin{align}
		x^{(k+2)} = x^{(k)} + {\varepsilon^2 \alpha} \big( \tfrac{\partial F}{\partial x}\big(x^{(k+1)}\big)  +   \tfrac{\partial F}{\partial x}\big(x^{(k)}\big)  \big) + \mathcal{O}(\varepsilon^3)
	\end{align}
	where $k \in \mathbb{N} $. With $ \xi_{1}^{(k)} := x^{(k)} $, $ \xi_{2}^{(k)} := x^{(k+1)} $ we can
	write this as
	\begin{align}
			\xi_{1}^{(k+1)} - \xi_{1}^{(k)} &= \xi_{2}^{(k)} - \xi_{1}^{(k)} \nonumber \\
			\xi_{2}^{(k+1)} - \xi_{2}^{(k)} &= \xi_{1}^{(k)} - \xi_{2}^{(k)} \\ 
			&+ {\varepsilon^2 \alpha} \big( \tfrac{\partial F}{\partial x}(\xi_{1}^{(k)}) + \tfrac{\partial F}{\partial x}(\xi_{2}^{(k)}) \big) + \mathcal{O}(\varepsilon^3). \nonumber
	\end{align}
	This is the Euler discretization (with step size $1$) of
	\begin{align}
		\begin{split}
			\dot{\xi}_1 &= \xi_2 - \xi_1 \\
			\dot{\xi}_2 &= \xi_{1} - \xi_{2} + \varepsilon^2 \alpha \big( \tfrac{\partial F}{\partial x}(\xi_{1}) + \tfrac{\partial F}{\partial x}(\xi_{2}) \big) + \mathcal{O}(\varepsilon^3).
		\end{split}
	\end{align}
	With $ z_1 := \xi_1 $, $ z_2 := \xi_2 - \xi_1 $, $ c_1 = c_2 = \varepsilon^2 \alpha $,
	$ \gamma = 1 $ and neglecting the $\mathcal{O}(\varepsilon^3)$-terms we arrive at \eqref{eqNewAlgorithm}.
	Thus, loosely speaking, \eqref{eqNewAlgorithm} can be interpreted as the averaged system 
	corresponding to the extremum seeking system discussed in \Cref{lemmaDeltaEqualEpsilon}.
	This relationship suggests that in certain cases extremum seeking algorithms do not mimic a simple gradient 
	flow but the more advanced method presented in \Cref{theoremNewAlgorithm}.
\end{remark}
\begin{remark}\label{remarkCombinationHeavyNesterov}
	The continuous-time algorithm \eqref{eqNewAlgorithm} can be interpreted as a combination of the continuous-time
	heavy ball method (see e.g. \cite{attouch2000heavy}) and a continuous-time version of Nesterov's 
	method as presented in \cite{duerr2012}. In the heavy ball method only the term $ \nabla F(z_1) $ 
	is present whereas in Nesterov's method only the term $ \nabla F(z_1+\gamma z_2) $ occurs.
	To be more precise, these are given by
	\begin{align}
		\dot{z}_1 = z_2 \qquad \dot{z}_2 = u
	\end{align}
	where $ u = -kz_2 -c_1\nabla F(z_1) $ for the heavy ball and $ u = -kz_2 -c_2\nabla F(z_1+\gamma z_2) $
	for Nesterov's method. Notice that if $ F$ is quadratic, i.e. the gradient is linear,
	there is no difference between Nesterov's method and \eqref{eqNewAlgorithm} up to the choice of parameters.
% 	the heavy ball method is given by
% 	\begin{align}
% 		\begin{split}
% 			\dot{z}_1 &= z_2 \\
% 			\dot{z}_2 &= -Kz_2 -c_1\nabla F(z_1)
% 		\end{split} \label{eqHeavyBall}
% 	\end{align}
% 	and Nesterov's method is
% 	\begin{align}
% 		\begin{split}
% 			\dot{z}_1 &= z_2 \\
% 			\dot{z}_2 &= -Kz_2 -c_2\nabla F(z_1+\gamma z_2).
% 		\end{split} \label{eqNesterov}
% 	\end{align}
	
\end{remark}
\begin{remark}
	The algorithm \eqref{eqNewAlgorithm} uses the gradient at two different points which
	can be interpreted as a simple way of averaging. Moreover, the term $ \nabla F(z_1 + \gamma \dot{z}_1) $
	uses knowledge about the derivative of $ z_1 $ which implements some kind of preview.
	For small $ \dot{z}_1 $, it is also $ \nabla F(z_1 + \gamma \dot{z}_1) \approx \nabla F(z_1) + \gamma \nabla^2 F(z_1) \dot{z}_1 $
	such that, close to the critical point, this term may also be interpreted as a curvature dependent damping.
\end{remark}
\myComment{Is it possible to investigate the new algorithm in discrete time using the results
of the Lessard-Paper? Can it then be compared to the heavy ball and Nesterov's method? How 
is the combination to be interpreted in terms of filtering?}
%

%%%%%%%%%%%%%%%%%%%%%%%%%%%%%%%%%%%%%%%%%%%%%%%%%%%%%%%%%%%%%%%%%%%%%%%%%%%%%%%%
\section{Example}\label{secExample}
In the following we illustrate the results from \Cref{theoremTwoNeedles} and \Cref{theoremNewAlgorithm}
by means of an example and simulations.
\begin{figure*}
	\def\figurewidth{0.44\textwidth}
	\def\figureheight{0.3\textwidth}
	\input{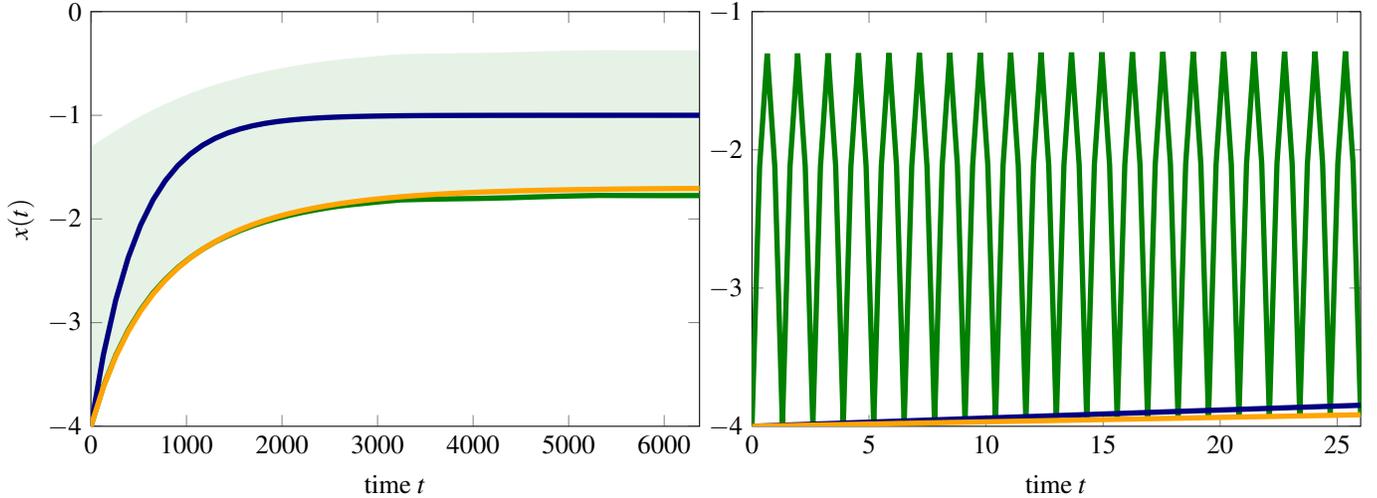}
	\caption{A comparison between the original system (green), i.e. system \eqref{eqNonlinearInputAffineSys} with inputs \eqref{eqInputSequence1_u1} and \eqref{eqInputSequence1_u2}, the approximation \eqref{eqApproximationTwoNeedles1} (orange)
 	         and a gradient descent algorithm (blue) in the case of quadratic $ F(x) $ ($b=2$,$c=3$,$\varepsilon=0.00001$,$T=1.3$,$x_{min}=-1$,$\alpha=-10$).
 	         On the left hand side the thick green line depicts the original system evaluated at multiples of the period length;
 	         the shaded area shows the original system jumping up and down very fast. A zoomed plot that illustrates this jumping
 	         can be seen on the right hand side.}\vspace*{-1.4em}
 	\label{figSimulationResultsQuadraticCase}
\end{figure*}
\subsection{The case of quadratic $F(x)$}
Suppose that $ F $ in \eqref{eqNonlinearInputAffineSys} is a scalar, quadratic function, i.e.
\begin{align}
	F(x) = x^2 + bx + c,   \label{eqQuadCaseF}
\end{align}
where $ b,c \in \mathbb{R} $ and $ F: \mathbb{R} \to \mathbb{R} $. The quadratic case
is important since many convex functions are locally quadratic in a region around their minimum.
For the input sequence defined by \eqref{eqInputSequence1_u1} and \eqref{eqInputSequence1_u2} one can compute
$ x^*(t) $ and $ \Phi(t,t_0) $ in \eqref{eqApproximationTwoNeedles1} analytically.
In particular, in case of $ 4c - b^2 \neq 0 $ one obtains
\begin{align}
	x^*(t) = \tfrac{1}{2} \big( \tan\big( p (t+K_j(x_j) ) \big) \sqrt{4c-b^2} - b \big) \label{eqQuadCaseSol}
\end{align}
and
\begin{align}
	\Phi(t,t_j) = \big\vert \frac{ \cos\big( p(t_j+K_j(x_j) ) \big) }{ \cos\big(p(t+K_j(x_j) )\big) }  \big\vert^{2} \label{eqQuadCaseTransitionMatrix}
\end{align}
for $ t \in [jT + \varepsilon,jT - \varepsilon+\tfrac{T}{2}] $ with
%$ p := \tfrac{1}{2} \sqrt{4c-b^2} $ and $ K_j(x_j) := - jT - \varepsilon + \tfrac{1}{p} \arctan( \tfrac{b+2x_j}{\sqrt{4c-b^2}} ) $.
\begin{align}
	p &:= \tfrac{1}{2} \sqrt{4c-b^2} \label{eqQuadCaseDefP} \\
	K_j(x_j) &:= - jT - \varepsilon + \tfrac{1}{p} \arctan( \tfrac{b+2x_j}{\sqrt{4c-b^2}} ). \label{eqQuadCaseDefK}
\end{align}
A derivation of these equations is given in the \Cref{secAppendix}.
Notice that this also includes the case of $ 4c-b^2 < 0 $, i.e. $ p $ is a complex
number, by the definition of the trigonometric functions for complex arguments but we
do not discuss this in detail here. 
The case of $ 4c - b^2 = 0 $ can be handled similarly and is not treated here as well.
% In case of $ 4c - b^2 = 0 $ we arrive at
% \begin{align}
% 	x^*(t) = - \frac{x_j+\tfrac{1}{2}b}{(t-jT-\varepsilon)(x_j+\tfrac{1}{2}b) - 1} -  \frac{1}{2} b \label{eqQuadCaseSol2}
% \end{align}
% and
% \begin{align}
% 	\Phi(t,t_j) = \frac{ (t_j-jT-\varepsilon)(x_j+\tfrac{1}{2}b) - 1 }{ (t-jT-\varepsilon)(x_j+\tfrac{1}{2}b) - 1 } \label{eqQuadCaseTransitionMatrix2}
% \end{align}
% for $ t \in [jT + \varepsilon,jT + \varepsilon+\tfrac{T}{2}] $.
Notice further that $ x^*(t) $ has finite escape time at
$
	t_{esc,j} = \tfrac{2m+1}{2p} \pi - K_j(x_j) 
$
with $ m \in \mathbb{Z} $.
% and in the case of $ 4c-b^2 $ it is at
% \begin{align}
% 	t_{esc,j} = \frac{2}{2x_j+b} + jT + \varepsilon .
% \end{align}
However, in most cases this is no problem in the implementation of the
approximation from \Cref{theoremTwoNeedles}. We implemented the iteration
\begin{align}
	x\big((j+1)T\big) &= x(jT) \label{eqQuadCaseApproxIteration} \\
	                  &\phantom{} + \varepsilon \alpha \Phi(jT,jT+\varepsilon)\big( 1 - \Phi(jT+\varepsilon,jT+\tfrac{T}{2}-\varepsilon) \big) \nonumber
\end{align}
which is another representation of \eqref{eqTwoNeedlesIteration} and where we neglected
the higher order terms. Thus, if $ t_{esc,j} \neq jT $ and $ t_{esc,j} \neq jT + \varepsilon $,
the evaluation makes no problem. 
As shown in the \Cref{secAppendix}, the iteration \eqref{eqQuadCaseApproxIteration}
has fix points at
\begin{align}
	\bar{x}_{1/2} &= \tfrac{1}{2} \big( \sqrt{4c-b^2} \frac{ \cos(\tfrac{pT}{2}-2\varepsilon) \mp 1 }{ \sin(\tfrac{pT}{2}-2\varepsilon) } - b \big) . \label{eqQaudCaseFixpoints}
\end{align}
Notice that the minimum of the function $ F = x^2 + bx + c $ %as defined in \eqref{eqQuadCaseF}
is at $ x_{min} = -\tfrac{b}{2} $. Thus, since $ \cos(\tfrac{pT}{2}-2\varepsilon) \mp 1 \neq 0 $
for $ \sin(\tfrac{pT}{2}-2\varepsilon) \neq 0 $, it is $ \bar{x}_{1/2} \neq x_{min} $ such that 
the iteration never converges to the minimum of $ F $. However, we can get arbitrarily
close.
Secondly, it is worth to mention that \eqref{eqQuadCaseApproxIteration} possesses
two fix points. Simulations suggest that one is asymptotically stable and the other
one is unstable where the stability depends on the chosen period length $ T $. \\
In our simulations we compared the approximative iteration \eqref{eqQuadCaseApproxIteration}
with a fixed step simulation of the original system, i.e. \eqref{eqNonlinearInputAffineSys} 
with inputs \eqref{eqInputSequence1_u1} and \eqref{eqInputSequence1_u2}, and a simple 
gradient descent. The simulation results are depicted in \Cref{figSimulationResultsQuadraticCase}.
At the respective time points $ jT $ our approximation is much closer to the original
system as the gradient descent which seems to correspond to an average of the original
system. This is to be expected since here the period length is large compared to the
length of the needle which is also the reason for the large jumps to be seen on the right hand side of \Cref{figSimulationResultsQuadraticCase}. 
In the considered example our approximation seems to be a good measure for the lower ``boundary'' of the trajectory
of the original system. Notice that also an approximation of the upper boundary can be computed similarly.
% \begin{figure}
% 	\includegraphics[width=\columnwidth]{figures/comparison3AndAtT_Cropped.pdf}
% 	\caption{A comparison between the original system (green), the approximation \eqref{eqApproximationTwoNeedles1} (red)
% 	         and a gradient descent algorithm (blue) in the case of quadratic $ F(x) $ ($b=2$,$c=3$,$\varepsilon=0.0001$,$T=1.3$).
% 	         The minimum is attained at $ x_{min}=-1$.}
% 	\label{figSimulationResultsQuadraticCase}
% \end{figure}
% \begin{figure}
% 	\includegraphics[width=\columnwidth]{figures/comparison3_ShortTime_Cropped.pdf}
% 	\caption{ff}
% 	\label{fiComparisonAtT}
% \end{figure}

\subsection{Simulative analysis of the algorithm \eqref{eqNewAlgorithm}}
\begin{figure}[t]
	\def\figurewidth{\columnwidth}
	\input{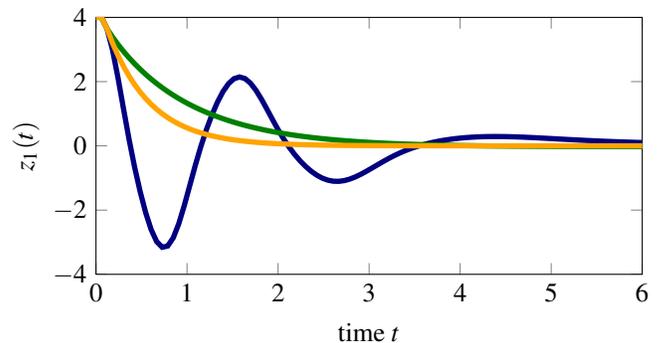}
	\caption{Comparison of the heavy ball method (blue), Nesterov's method (green) and the proposed
	         algorithm \eqref{eqNewAlgorithm} (orange) for the minimization of $ F(x) = \vert x \vert^3 $.
	         The parameters of the new algorithm are chosen as $ c_1 = c_2 = \gamma = 1 $, $ K = 2 $.
	         For Nesterov's method they are chosen such that for 
	         the case of quadratic $ F(x) $ it is equal to the proposed algorithm.
	         The numerical integration is done using a standard fixed step Euler method.}
	\label{figComparisonAlgorithms}
\end{figure}

We compared the proposed algorithm \eqref{eqNewAlgorithm} to the heavy ball
method and a continuous-time version of Nesterov's method (see also
\Cref{remarkCombinationHeavyNesterov}). Simulation results for the case 
of $ F(x) = \vert x \vert^3 $ are depicted in \Cref{figComparisonAlgorithms}.
In the considered case the proposed algorithm shows a fast convergence
without overshoot in comparison to both other algorithms. Further simulations
with different objective functions or varied parameters show a similar
behavior. However, the simulation results should be interpreted with care
when it comes to performance or convergence speed since all three algorithms
include parameters and it is not clear how to choose them in a way such 
that direct comparisons are possible.

%%%%%%%%%%%%%%%%%%%%%%%%%%%%%%%%%%%%%%%%%%%%%%%%%%%%%%%%%%%%%%%%%%%%%%%%%%%%%%%%
\section{Conclusions and outlook}\label{secConclusions}
In this work we introduced needle-shaped dither signals for gradient approximation
and extremum seeking. We derived formulas that give 
insight into the averaging process of extremum seeking schemes with
needle-shaped dither signals. We further showed how this can be generalized to 
arbitrary periodic dither signals by superposition. Thus, needle-shaped
dithers can be seen as basis functions for a wide range of more general signals.
Motivated by these results we also proposed a new gradient-based optimization algorithm. The algorithm
is related to well-known accelerated gradient methods and is of interest on its own. 
By taking two gradients into account the behavior of the proposed algorithm is
similar to the averaged behavior of the extremum seeking scheme with needle-shaped dither signals.
This might be one hint why extremum seeking schemes often perform relatively well
in practice despite their simplicity. \\
Since our approach relies on well-established
ideas from the Maximum Principle we hope that we can extend our setup using existing generalizations 
of the Maximum Principle (\cite{sussmann2002needle}). As already mentioned before,
we also expect that our results can be extended to the multidimensional case.
Moreover, we aim to use our new knowledge about the gradient approximation process to
design dither signals that are in some terms optimal.

%%%%%%%%%%%%%%%%%%%%%%%%%%%%%%%%%%%%%%%%%%%%%%%%%%%%%%%%%%%%%%%%%%%%%%%%%%%%%%%%
\setstretch{1}
\bibliographystyle{plain}
\bibliography{bibfile}

%%%%%%%%%%%%%%%%%%%%%%%%%%%%%%%%%%%%%%%%%%%%%%%%%%%%%%%%%%%%%%%%%%%%%%%%%%%%%%%%
\section{Appendix}\label{secAppendix}
\subsection{Derivation of \eqref{eqQuadCaseSol}, \eqref{eqQuadCaseTransitionMatrix}} %, \eqref{eqQuadCaseSol2} and \eqref{eqQuadCaseTransitionMatrix2}}
For $F$ as defined by \eqref{eqQuadCaseF} and $ t \in [jT + \varepsilon,jT - \varepsilon+\tfrac{T}{2}]$, 
$ j \in \mathbb{N} $,  $ x^*(t) $ is the solution of
\begin{align}	
	\dot{x}^*(t) = {x^*}^2(t) + b x^*(t) + c, \qquad x(jT+\varepsilon) =: x_j.
\end{align}
We will now solve this differential equation via separation of variables. Resorting and
integrating gives
\begin{align}
	\int_{x_j}^{x^*} \tfrac{1}{\xi^2+b\xi+c} \, d\xi &= \int_{jT+\varepsilon}^t 1 \, d\tau.
\end{align}
These integrals can be solved using standard formulas for elementary functions, see e.g. 
\cite{abramowitz1972handbook}. If $ 4c - b^2 \neq 0 $ we have that
\begin{align}
	\bigg[ \tfrac{2}{\sqrt{4c-b^2}} \arctan( \tfrac{b+2\xi}{\sqrt{4c-b^2}} ) \bigg]_{x_j}^{x^*} &= t - jT - \varepsilon .
\end{align}
Notice that this also includes the case of $ 4c-b^2 < 0 $ by the definition of
the $ \arctan $ as 
$
	\arctan(z) = \tfrac{i}{2} \ln( \tfrac{i+z}{i-z} ) 
$
(see \cite{abramowitz1972handbook})
where $ i = \sqrt{-1} $ is the imaginary unit. Thus, solving this equation for $ x^* $ we obtain the solution as given by \eqref{eqQuadCaseSol}.
% In case of $ 4c - b^2 = 0 $, it is $ \xi^2+b\xi+c = ( \xi + \tfrac{1}{2}b)^2 $ and hence we obtain
% \begin{align*}
% 	\bigg[ -\frac{1}{\xi + \frac{1}{2}b} \bigg]_{x_j}^{x^*} &= t - jT - \varepsilon  .
% \end{align*}
% Again, solving for $ x^* $, we arrive at \eqref{eqQuadCaseSol2}.
To obtain the transition matrix $ \Phi(t,t_j) $ we use that the variational equation
\eqref{eqTwoNeedlesVariationalEquation} here is a scalar linear time varying differential
equation such that (see e.g. \cite{kailath1980linear})
$
	\Phi(t,t_j) = \exp\big( \int_{t_j}^{t} \tfrac{\partial F}{\partial x}(x^*(\tau)) \, d\tau \big).
$
Putting the solution \eqref{eqQuadCaseSol} into this equation we obtain
\begin{align}
	\Phi(t,t_j) &= \exp\bigg( \int_{t_j}^{t} \sqrt{4c-b^2} \tan\big( p (\tau+K_j(x_j)) \big) d\tau \bigg) \nonumber \\
	            &= \exp\bigg( \sqrt{4c-b^2} \bigg[ - \frac{ \ln\big( \big\vert \cos(p(\tau+K_j(x_j))) \big\vert \big) }{p} \bigg]_{t_j}^{t} \bigg) \nonumber \\
	            %&= \exp\bigg( - \tfrac{\sqrt{4c-b^2}}{p} \ln\big( \big\vert \frac{\cos(p(t+K_j(x_j)))}{\cos(p(t_j+K_j(x_j)))} \big\vert \big)  \bigg) \nonumber \\
	            &= \big\vert \frac{ \cos\big( p(t+K_j(x_j)) \big) }{ \cos\big(p(t_j+K_j(x_j))\big) }  \big\vert^{-2}.
\end{align}
%\Cref{figPlotPhi} shows a plot of $ \Phi(t,t_j) $.
% Equivalently, with \eqref{eqQuadCaseSol2} in the case of $ 4c - b^2 = 0 $, we obtain
% \begin{align*}
% 	\Phi(t,t_j) &= \exp\bigg( \int_{t_j}^{t} - \frac{2x_j+b}{(\tau-jT-\varepsilon)(x_j+\tfrac{1}{2}b) - 1} d\tau \bigg) \\
% 	            &= \exp\bigg( - \bigg[ \ln \big( (\tau-jT-\varepsilon)(x_j+\tfrac{1}{2}b) - 1 \big) \bigg]_{t_j}^{t} \bigg) \\
% 	            &= \exp\bigg( \ln( \frac{ (t_j-jT-\varepsilon)(x_j+\tfrac{1}{2}b) - 1 }{ (t-jT-\varepsilon)(x_j+\tfrac{1}{2}b) - 1 } ) \bigg) \\
% 	            &= \frac{ (t_j-jT-\varepsilon)(x_j+\tfrac{1}{2}b) - 1 }{ (t-jT-\varepsilon)(x_j+\tfrac{1}{2}b) - 1 }.
% \end{align*}
% \def\figurewidth{\columnwidth}
% \def\figureheight{4.3cm}
% \begin{figure}
% 	\input{figures/plotPhi.tex}
% 	\caption{Plot of the state-transition matrix \eqref{eqQuadCaseTransitionMatrix} in different time intervals. 
% 	         The period in which $ \Phi(t,t_j) $ is evaluated in \eqref{eqApproximationTwoNeedles1} is marked in orange.}
% 	\label{figPlotPhi}
% \end{figure}

\subsection{Fix points of the iteration \eqref{eqQuadCaseApproxIteration}}
We briefly analyze the iteration \eqref{eqQuadCaseApproxIteration}. 
Let $ t_{j1} := jT+\varepsilon $ and $ t_{j2} := jT+\tfrac{T}{2}-\varepsilon $.
Since $ \varepsilon > 0 $, $ \alpha > 0 $, \eqref{eqQuadCaseApproxIteration}
has a fix point at $ \bar{x} $ implicitly given by the equation
\begin{align}
	1 - \Phi(t_{j1},t_{j2}) &= 1 - \big\vert \tfrac{ \cos\big( p(t_{j2}+K_j(\bar{x}) ) \big) }{ \cos\big(p(t_{j1}+K_j(\bar{x}) )\big) }  \big\vert^{2} 
	                                                    = 0.
\end{align}
By \eqref{eqQuadCaseTransitionMatrix} we have with \eqref{eqQuadCaseDefP} and \eqref{eqQuadCaseDefK}
\begin{align}
	\Phi(t_{j1},t_{j2}) &= \big\vert \tfrac{ \cos\big(\tfrac{pT}{2} - 2p \varepsilon + \arctan(\tfrac{b+2x_j}{\sqrt{4c-b^2}}) \big) }{ \cos\big(\arctan(\tfrac{b+2x_j}{\sqrt{4c-b^2}}) \big) } \big\vert^2 
\end{align}
and using the trigonometric identity 
\begin{align}
	\tfrac{\cos(\alpha+\beta)}{\cos(\beta)} = \cos(\alpha) + \sin(\alpha)\tfrac{\sin(\beta)}{\cos(\beta)} 
\end{align}
we obtain 
\begin{align}
	\Phi(t_{j1},t_{j2}) = \big\vert \cos(\alpha) - \sin(\alpha) \tfrac{b+2x_j}{\sqrt{4c-b^2}} \big\vert^2
\end{align}
with \mbox{$ \alpha := \tfrac{pT}{2}- 2p\varepsilon $}. 
We compute further
\begin{align}
	 &~ 1 - \Phi(t_{j1},t_{j2}) \nonumber \\
	=\:&~ 1 - \cos^2(\alpha) + 2 \sin(\alpha)\cos(\alpha) \tfrac{\alpha}{\sqrt{4c-b^2}}  + \sin^2(\alpha) \tfrac{(b+2x_j)^2}{4c-b^2} \nonumber \\
	=\:&~ \sin^2(\alpha) \big( 1 - \tfrac{(b+2x_j)^2}{4c-b^2} \big) + 2 \sin(\alpha)\cos(\alpha) \tfrac{b+2x_j}{\sqrt{4c-b^2}}.
\end{align}
In the following we assume that $ \sin(\alpha) \neq 0 $. Then, using the
previous results, the equation for the fix point is given by
\begin{align}
	 \sin(\alpha) \big( 1 - \tfrac{(b+2\bar{x})^2}{4c-b^2} \big) + 2 \cos(\alpha) \tfrac{b+2\bar{x}}{\sqrt{4c-b^2}} = 0 .
\end{align}
Let $ w:= \tfrac{(b+2\bar{x})}{\sqrt{4c-b^2}} $. Then this is a quadratic equation
in $ w $ and its solutions are given by
$
	w_{1/2} %&= \frac{ -2\cos(\alpha) \pm \sqrt{ 4 \cos^2(\alpha) + 4 \sin^2(\alpha) } }{-2\sin(\alpha)} \nonumber \\
	        = \frac{ \cos(\alpha) \mp 1 }{ \sin(\alpha) } .
$
Thus, with $ \alpha = \tfrac{pT}{2}- 2p\varepsilon $, the fix points are given as in \eqref{eqQaudCaseFixpoints}.

\end{document}